\documentclass[11pt,twoside,a4paper]{article}
\usepackage{amsmath}
\usepackage{amssymb}
\usepackage{amsthm}
\usepackage{cite}
\usepackage{verbatim}
\usepackage{fullpage}
\usepackage{graphicx}
\usepackage{epstopdf}
\usepackage{caption}
\usepackage{subcaption}
\usepackage{color}
\makeatletter
\def\Ddots{\mathinner{\mkern1mu\raise\p@
\vbox{\kern7\p@\hbox{.}}\mkern2mu
\raise4\p@\hbox{.}\mkern2mu\raise7\p@\hbox{.}\mkern1mu}}
\graphicspath{{C:/Users/Alexander/Dropbox/Vortices_and_Impurities/Vortices_and_magnetic_impurities/}}
\makeatother
\numberwithin{equation}{section}
\title{Dynamics of vortices with magnetic impurities}

\author{Alexander Cockburn\thanks{alexander\_cockburn12@hotmail.com}~\\[5pt]
{\normalsize {\sl Department of Mathematical Sciences,}}\\
{\normalsize {\sl Durham University, Durham DH1 3LE,}}\\
{\normalsize {\sl United Kingdom,}}\\[5pt]
Steffen Krusch\thanks{S.Krusch@kent.ac.uk} and
Abera A. Muhamed\thanks{A.A.Muhamed@kent.ac.uk}~\\[5pt]
{\normalsize {\sl School of Mathematics, Statistics and Actuarial
Science}}\\
{\normalsize {\sl University of Kent,
Canterbury CT2 7NF}}\\
{\normalsize {\sl United Kingdom.}}
}

\begin{document}
{\let\newpage\relax\maketitle}

\begin{abstract}
We investigate the dynamics of BPS vortices in the presence of magnetic impurities taking the form of axially-symmetric localised lumps and delta-functions.  We present numerical results for vortices on flat space, as well as exact results for vortices on hyperbolic space in the presence of delta-function impurities.  In fact, delta-function impurities of appropriate strength can be captured within the moduli space approximation by keeping one or more of the vortices fixed.  We also show that previous work on vortices on the 2-sphere extends naturally to the inclusion of delta-function impurities.
\end{abstract}

\section{Introduction}
In a recent paper \cite{TW14} Tong and Wong discussed BPS vortices in the presence of both electric and magnetic impurities.  At the classical level, electric impurities were shown to leave the moduli space of static solutions unchanged, but the usual geodesic approximation to the dynamics is supplemented by a connection term.  This observation allowed an analysis of both the classical and quantum dynamics of a single vortex in the presence of a delta-function electric impurity. The purpose of this paper is to investigate the less well-understood effect of magnetic impurities on vortex dynamics.  

The paper \cite{TW14} has sparked some recent interest in existence results for vortices in certain product Abelian gauge theories, which are related to vortices with magnetic impurities by taking the mass of one of the vortex species to infinity. Sharp existence theorems for static solutions in such models have been proven in \cite{Zhang}. General existence results allowing the coexistence of vortices and anti-vortices subject to certain inequalities have been obtained in \cite{Han:2015yua}, and similar results hold for a related Abelian 
Chern-Simons-Higgs model \cite{Han:2015tga}.  This paper complements these existence results and the original arguments given in \cite{TW14} with some numerical evidence, but our main interest is in the vortex dynamics.  To illustrate the generic effect of magnetic impurities on the dynamics, we give numerical examples of moduli space metrics for a single vortex in the presence of axially-symmetric, localised lump-like impurities.  We also investigate the limit where the impurity approaches a delta-function, and present numerical evidence that this limit is well-defined.

While no explicit solutions for vortices on flat space are known, progress can be made by changing the background.  Witten observed \cite{W78} that the vortex equations on the hyperbolic plane are integrable for a particular value of the curvature, and this allowed an explicit calculation of the 2-vortex moduli space metric by Strachan \cite{S92}.  Baptista and Manton gave analytic approximations for vortex solutions and moduli space metrics on the 2-sphere in the limit where the volume of the sphere is just big enough to accommodate the vortices \cite{BM03}.  This paper adapts the results on these different backgrounds to the inclusion of a delta-function magnetic impurity, showing that explicit expressions are still obtainable.

The paper is structured as follows. In section 2 we review vortices in
flat space, discuss the effects of magnetic impurities, and present
numerical results. In section 3 we consider vortices with impurities on
hyperbolic space, describe the corresponding moduli space and derive the
moduli space metric explicitly. In section 4 we discuss vortices on the
2-sphere. We derive the metric for one vortex in the presence of an
impurity near the Bradlow limit and discuss its dynamics. We end with a conclusion.

\section{Flat space vortices with magnetic impurities}
\subsection{Previous work on impurity-free flat space vortices}
We begin by reviewing relevant previous work on flat-space vortices.  The standard action for vortices at critical coupling is:
\begin{equation}\label{abhiggs}
\int\left( -\frac{1}{4}F_{\mu\nu}F^{\mu\nu}+\frac{1}{2}\overline{D_\mu\phi}D^\mu\phi
-\frac{1}{8}(1-|\phi|^2)^2 \right)d^3x,
\end{equation}
where $\phi$ is a complex scalar field coupled to a $U(1)$ gauge field $A_\mu$, $D_\mu\phi=(\partial_\mu-iA_\mu)\phi$ is the covariant derivative and the gauge field strength is $F_{\mu\nu}=\partial_\mu A_\nu-\partial_\nu A_\mu$.  The signature of the metric is taken to be $(1,-1,-1)$.  After imposing the equation of motion associated with the gauge choice $A_0=0$ as a constraint, the Lagrangian can be written in terms of kinetic and potential energies as $L=T-V$, where (for $i=1,2$):
\begin{align}
T&=\frac{1}{2}\int \left(|\dot{\phi}|^2+\dot{A_i}\dot{A_i}\right)d^2x, \label{kinetic}\\
V&=\frac{1}{2}\int \left(\overline{D_i\phi}D_i\phi + B^2 + \frac{1}{4}\left(1-|\phi|^2\right)^2\right)d^2x,
\end{align}
and the total conserved energy is $E=T+V$.  Suppose now that the fields are static, so that $\dot{\phi}=\dot{A}=0$, and $T$ vanishes.  We can apply a standard Bogomolny argument to the potential energy:
\[
V=\frac{1}{2}\int\left((D_1\pm iD_2)\phi\overline{(D_1\pm iD_2)\phi}+\left(B\mp\frac{1}{2}(1-|\phi|^2)\right)^2\pm B\right)d^2x,
\]
implying the topological bound
\[
E\geq \left\lvert\frac{1}{2} \int B\, d^2x\right\rvert=\pi |N|,
\]
where $N$ is the degree of the map $\phi|_{|x|\to\infty}:S^1\to U(1)$.  For $N>0$, this bound is saturated when the following first-order Bogomolny equations are satisfied:
\begin{align}
D_1\phi+ iD_2\phi&=0,\label{bog1}\\
B- \frac{1}{2}(1-|\phi|^2)&=0.\label{bog2}
\end{align}
The sign of $N$ is reversed by a reflection, so we shall only consider the $N>0$ case.  Although no non-trivial explicit solutions to \eqref{bog1} and \eqref{bog2} are known, Taubes proved \cite{JT80} that given $N$ points $\{z_r\}_{1\leq r\leq N}$ on $\mathbf{R}^2$ there exists a solution for which $\phi$ vanishes precisely at the points $z_r$, and this solution is unique up to gauge equivalence.  Taubes also made the important observation that we can rewrite \eqref{bog1} and \eqref{bog2} as a single equation.  If we define the gauge-invariant quantity $f=\log |\phi|^2$, then we can use \eqref{bog1} to solve for $A_\mu$ in terms of $\phi$ and substitute this expression into \eqref{bog2} to obtain
\begin{equation}\label{originalTaubes}
\nabla^2 f+1-e^f=4\pi\sum^N_{r=1}\delta^{2}(z-z_r).
\end{equation}

The dynamics of slow-moving BPS solitons is captured by the natural metric on the moduli space of gauge-equivalent static solutions \cite{M81}.  In this approximation, the Lagrangian is given by the kinetic energy integral \eqref{kinetic} restricted to the moduli space.  This means that $(\dot{\phi},\dot{A}_i)$ is taken to satisfy the linearised version of the Bogomolny equations, as well as the linearised version of Gauss's law to ensure that $(\dot{\phi},\dot{A}_i)$ is orthogonal to the gauge orbit through $(\phi,A_i)$.  In other words, this approximation models the dynamics as a sigma model with the moduli space as target space, and the dynamics correspond to free geodesic motion on the moduli space.  This approximation has been proved rigorously for both vortices \cite{Stu94vor} and monopoles \cite{Stu94mon} by Stuart.  We refer the reader to \cite{MS04} for more details on the moduli space approximation.

This moduli space metric for vortices was explored in detail by Samols \cite{S91}, who showed that the kinetic energy integral localises around the vortex zeroes.  If we expand $f$ around a zero $z_r$ as
\begin{equation}\label{fexpansion}
f=\log |z-z_r|^2+a_r+\frac{1}{2}\left( b_r(z-z_r)+\bar{b}_r(\bar{z}-\bar{z}_r)\right)+c_r(z-z_r)^2+d_r(z-z_r)(\bar{z}-\bar{z}_r)+\bar{c}_r(\bar{z}-\bar{z}_r)^2+\dots
\end{equation}
then Samols' formula for the metric is
\begin{equation}\label{Sformula}
ds^2=\pi\sum^N_{r,s=1}\left( \delta_{rs}+2\frac{\partial \bar{b}_s}{\partial z_r}\right)dz_r\, d\bar{z}_s.
\end{equation}

\subsection{Flat space vortices with impurities}\label{flatimps}
The deformation of the action \eqref{abhiggs} suggested in \cite{TW14} to include magnetic impurities is:
\begin{equation}\label{mag}
\int \left(-\frac{1}{4}F_{\mu\nu}F^{\mu\nu}+\frac{1}{2}\overline{D_\mu\phi}D^\mu\phi
-\frac{1}{8}(1+\sigma-|\phi|^2)^2+\frac{1}{2}\sigma B\right)\, d^3x,
\end{equation}
where $\sigma$ is a static source for the magnetic field $B=F_{12}$.  In order for the static energy to remain real and finite, we take $\sigma$ to be real-valued with finite $L^2$ norm.  Applying the usual Bogomolny argument gives first-order equations:
\begin{align}
D_1\phi+iD_2\phi&=0,\label{modbog1}\\
B-\frac{1}{2}(1+\sigma-|\phi|^2)&=0,\label{modbog2}
\end{align}
and we still have the topological bound $E\geq\frac{1}{2}\int B\, d^2x=\pi N$, where $N$ is the asymptotic winding of $\phi$.  Two arguments are given in \cite{TW14} for the existence of a $2N$-dimensional moduli space of solutions to these equations.  Firstly, the linearisation of equations \eqref{modbog1} and \eqref{modbog2} and Gauss's law is independent of the impurity $\sigma$, so the usual index theorem \cite{W79a} counting the number of linearised deformations goes through unchanged.  Secondly, the magnetic impurities above can be shown to arise as the limit of vortices in product gauge groups, and these systems can be realised as D-brane configurations.  More recently Zhang has extended these arguments by proving sharp existence theorems for vortices in product gauge groups \cite{Zhang}. The aim of this section is to adapt the techniques of the previous one to obtain numerical solutions for the case where $\sigma$ is an axially-symmetric localised impurity of the form $\sigma=ce^{-d(x^2+y^2)}$, where $c\in\mathbf{R}$ and $d\in\mathbf{R}^+$, and to investigate their qualitative behaviour. 

We can straightforwardly rewrite \eqref{modbog1} and \eqref{modbog2} as a modified version of Taubes' equation:
\begin{equation}\label{Taubes}
\nabla^2 f+1+\sigma-e^f=4\pi\sum^N_{r=1}\delta^{2}(z-z_r).
\end{equation}
The function $f$ has singularities, so for numerical work we solve for the function
\[
\Phi=f-\sum^N_{r=1}\log |z-z_r|^2,
\]
so that \eqref{Taubes} becomes
\begin{equation}\label{modTaubes}
\nabla^2 \Phi+1+\sigma-\prod^N_{r=1}|z-z_r|^2 e^\Phi=0,
\end{equation}
with boundary conditions
\begin{equation}\label{boundary}
\Phi\sim -\sum^N_{r=1}\log |z-z_r|^2\text{ as }|z|\to \infty.
\end{equation}

\begin{figure}[h]
\begin{center}
\includegraphics[scale=0.9]{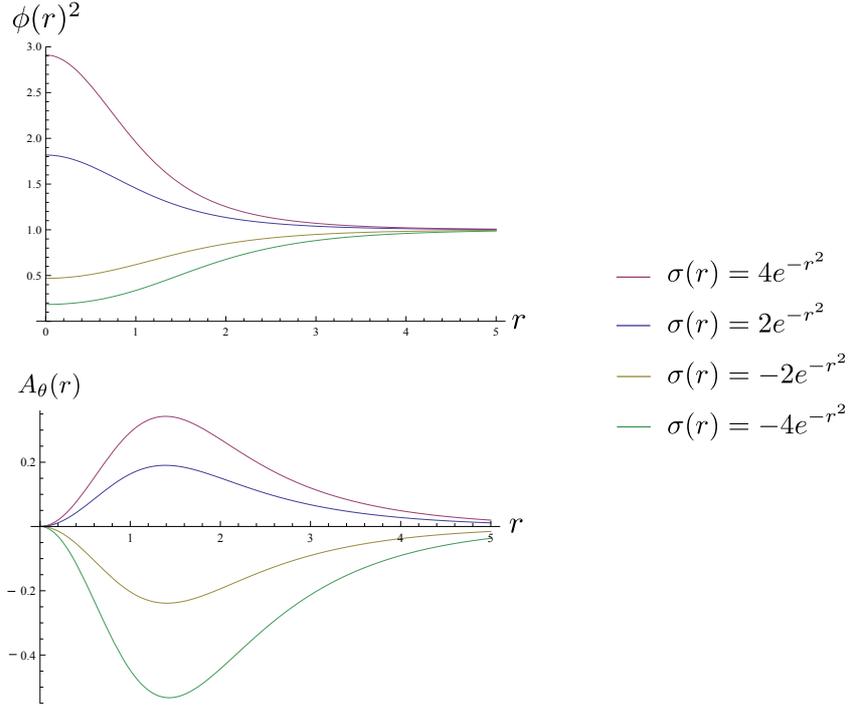}
\end{center}
\footnotesize{\caption{Higgs and gauge field profiles for vacua in the presence of various impurities.\label{vacimps}}}
\end{figure}

The first step is to solve for the vacuum. Plots of $\phi, A_\theta$ are given below for various values of $c,d$ in Fig.~\ref{vacimps}.  For these plots we have chosen an axial gauge where $A_r=0$ and $\phi$ is real on the whole plane, which is only possible for these vacuum solutions with zero asymptotic winding. The solutions were found using an over-relaxation method on the interval $[0,5]$ by imposing the Neumann boundary condition that $\Phi^\prime (0)=0$ and the Dirichlet condition \eqref{boundary} at $r=5$.  The solutions illustrate the important fact that the response of the fields to a localised impurity is also localised.  These plots also indicate that $\phi(0)\to 0$ as $c\to -\infty$, while $\phi(0)\to \infty$ as $c\to \infty$, and this limiting behaviour has been checked numerically over a much greater range of values of $c$ than those presented here.

\begin{figure}[h]
\begin{center}
\includegraphics[scale=0.85]{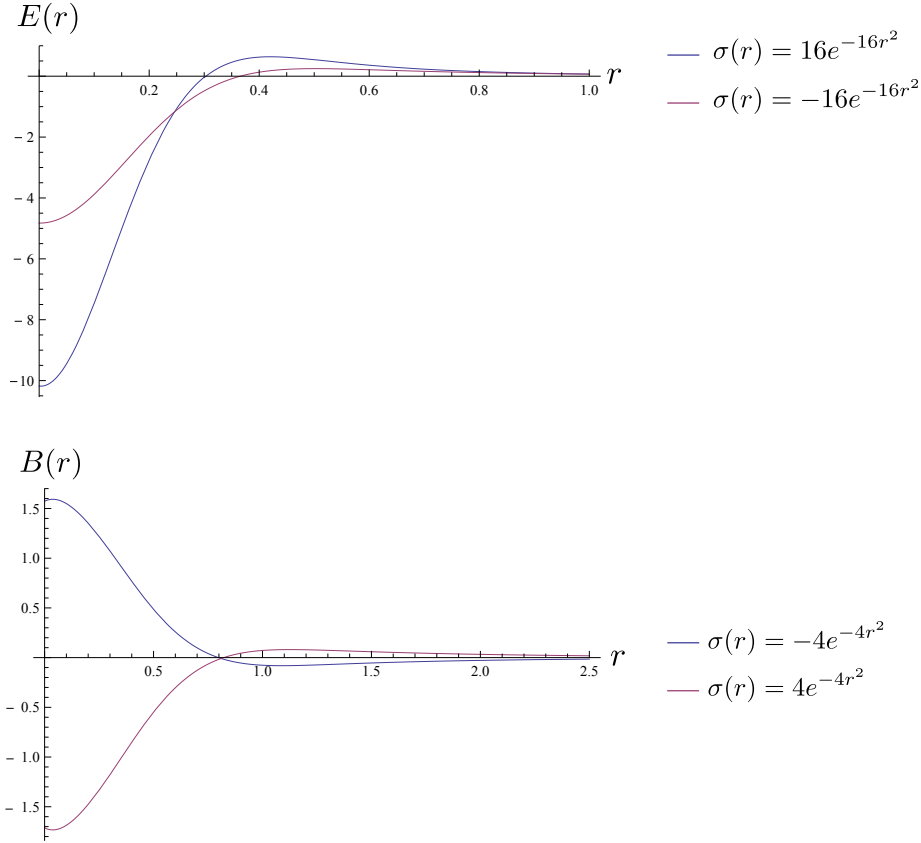}
\end{center}
\footnotesize{\caption{Energy densities and magnetic fields for vacua in the presence of impurities.\label{EB}}}
\end{figure}

Energy densities and magnetic fields are shown in Fig.~\ref{EB}.  Note that there is a range of values for which the energy density is negative.  In contrast to the standard abelian-Higgs system \eqref{abhiggs}, the integrand of the potential energy functional for the system with magnetic impurities is not a sum of total squares, so there is nothing to prevent this.  The plot of the magnetic field for the vacuum solution illustrates the general observation that reversing the sign of the impurity appears to approximately reverse the sign of $B$.

\begin{figure}[h]
\begin{center}
\includegraphics[scale=0.87]{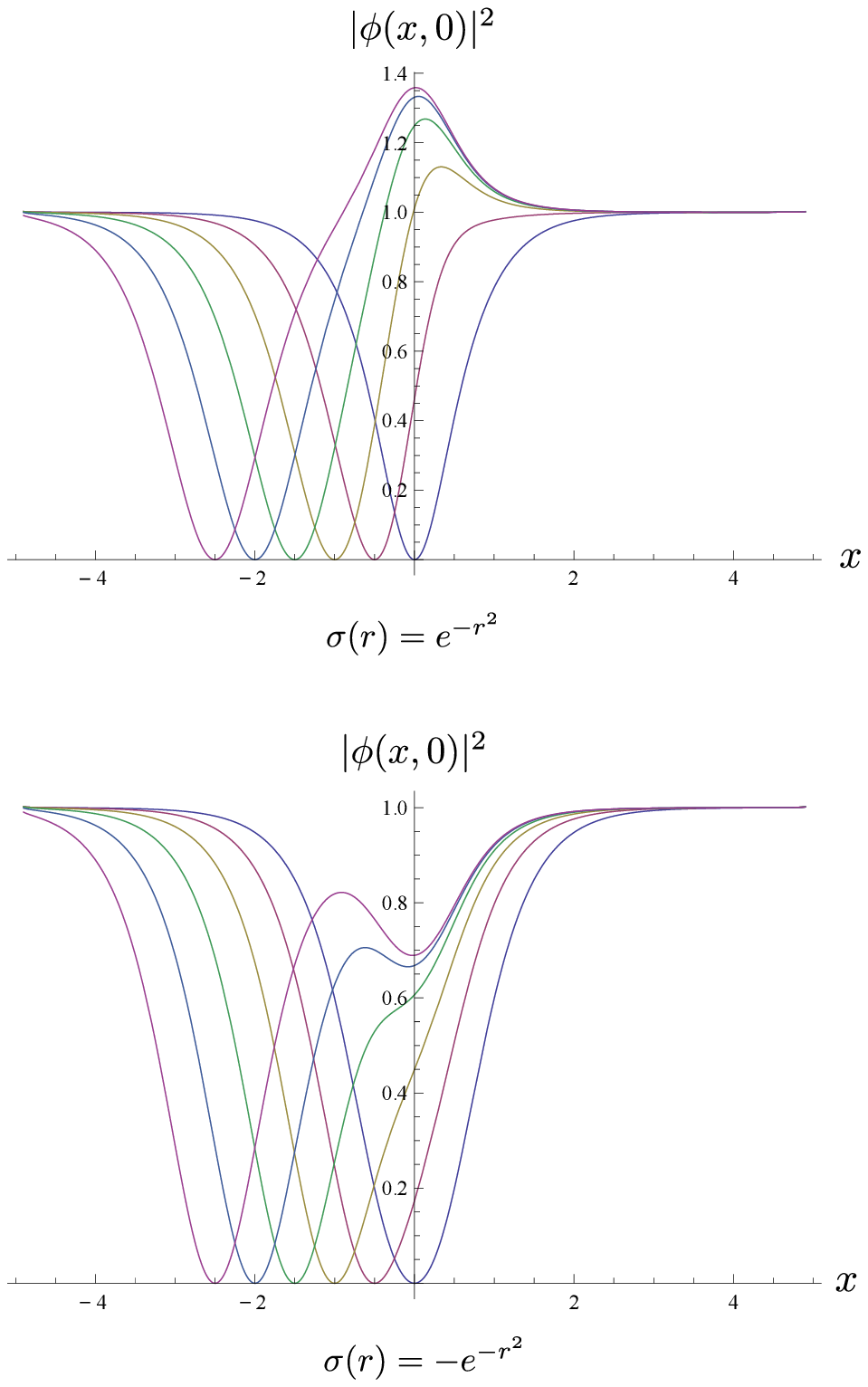}
\end{center}
\footnotesize{\caption{Higgs fields for families of 1-vortices in the presence of two different axial impurities.\label{1vortices}}}
\end{figure}

To find 1-vortex solutions, we can numerically solve \eqref{modTaubes} with $N=1$ by over-relaxing from an initial configuration 
\[
\Phi_0=\log (\rho(|z-z_1|))^2-\log |z-z_1|^2+\Phi_{\text{vac}},
\]
where we have approximated the 1-vortex profile by $\rho(r)=\tanh(0.6r)$ and $\Phi_{\text{vac}}$ is one of the vacuum-impurity solutions found above. Some solutions for $\sigma(r)=\pm e^{-r^2}$ in the presence of a vortex at $z_1$ are plotted in Fig.~\ref{1vortices} with $z_1$ taking values on the $x$-axis between $-2.5$ and $0$ in steps of 0.5.  The fact that solutions appear to exist wherever one puts the vortex zero provides evidence of a 1-vortex moduli space. The energies of these solutions are within 1\% of the Bogomolny bound, giving a good check on the numerics.  As one would expect, the solution looks like a superposition of the vacuum solution and an ordinary 1-vortex when the vortex zero is placed far from the impurity, but the vortex appears to `screen' the impurity as it approaches the origin.
\subsection{Moduli space metrics}
\begin{figure}[h]
\begin{center}
\includegraphics[scale=0.87]{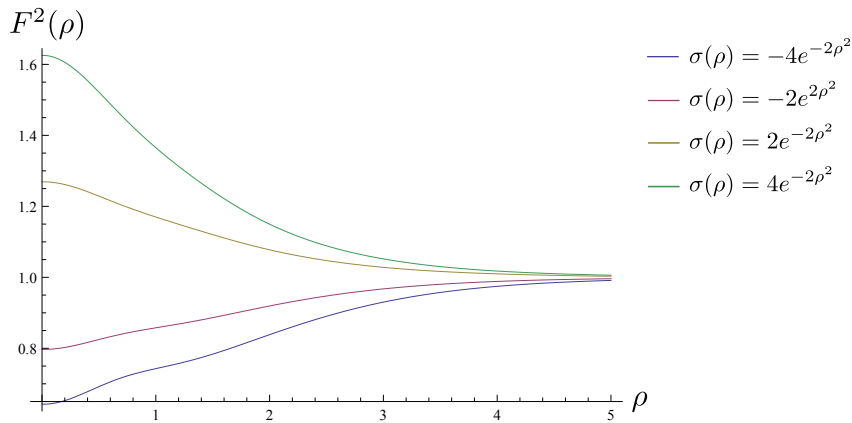}
\footnotesize{\caption{Moduli space metric profiles for 1-vortices in the presence of various impurities.\label{metrics}}}
\end{center} 
\end{figure}

Samols' formula \eqref{Sformula} generalises easily to include magnetic impurities.  The only difference is due to the fact the quantity $d_r$ defined in \eqref{fexpansion}, which in the impurity-free case is equal to $-1/4$, must be equal to $-\frac{1}{4}(1+\sigma(z_r))$ to satisfy \eqref{modTaubes}.  There are still no local constraints on $a_r$, $b_r$ and $c_r$, and the rest of the derivation goes through as in the impurity-free case \cite{S91}.  The general expression for the metric is:
\begin{equation}\label{modmetric}
ds^2=\pi\sum^N_{r,s=1}\left( \delta_{rs}(1+\sigma(z_r))+2\frac{\partial \bar{b}_s}{\partial z_r}\right)dz_r d\bar{z}_s.
\end{equation}
Our numerical metrics will be for 1-vortices in the presence of axially symmetric impurities.  If $z_1=\rho e^{i\theta}$ is the position of the vortex zero, then by rotational symmetry and the fact that the metric is Hermitian $b_1$ must take the form $b_1=b(\rho)e^{-i\theta}$, and the metric is
\begin{align}
ds^2&=\pi\left(1+\sigma(\rho)+2\frac{\partial \bar{b}_1}{\partial z_1}\right) dz_1 d\bar{z}_1\\
&=\pi\left(1+\sigma(\rho)+\frac{1}{\rho}\frac{d(\rho b)}{d\rho}\right)\left(d\rho^2+\rho^2d\theta^2\right)\label{impmetric}\\
&\equiv \pi F^2(\rho)\left(d\rho^2+\rho^2d\theta^2\right).\label{Fdefn}
\end{align}
For the 1-vortex, $b_1=2\partial_z\Phi(z_1)$, so $b=\partial_\rho\Phi(z_1)$. Fig.~\ref{metrics} illustrates the results of this for different impurities.  Numerically calculating geodesics for these is straightforward, and they show that in general a slow-moving vortex is repelled from the impurity if $c<0$, and attracted if $c>0$.  

We can obtain a more physical understanding of this behaviour using a point particle approximation of both the vortex and the impurity.  Just as for impurity-free vortices \cite{MS03}, this gives analytic information about the moduli space metric when the vortex is far from the impurity.  Suppose we have either an axial impurity on its own or a single vortex at the origin.  Provided the impurity decays sufficiently rapidly, then in either case \eqref{Taubes} linearises at large $\rho$ to
\begin{equation}\label{Bessel}
\frac{d^2f_0}{d\rho^2}+\frac{1}{\rho}\frac{df_0}{d\rho}-f_0=0,
\end{equation}
the modified Bessel equation of zeroth order, and this is independent of the impurity. $f_0$ must therefore have the asymptotic form
\[
f_0(\rho)\sim 2qK_0(\rho)
\]
for some constant $q$.  The interpretation is that at large $\rho$, the vortex or impurity can be thought of as a composite of a scalar monopole of charge $q$ and a magnetic dipole of moment $q$ perpendicular to the plane \cite{Sp96}.  We shall therefore refer to $q$ as the point charge of the vortex or impurity.  Manton and Speight used this point particle model to calculate the asymptotic $N$-vortex moduli space metric in \cite{MS03}.

For a 1-vortex, the point charge was first calculated numerically by de Vega and Schaposnik \cite{deVega:1976mi} to be $1.7079$, and later Tong gave a string theory argument suggesting that the point charge is $-8^\frac{1}{4}$ \cite{T02}.  In a detailed numerical study Ohashi showed in \cite{Ohashi:2015yta} that while being remarkably accurate Tong's value slightly underestimates the true value.
The point charges of impurities are straightforward to calculate numerically:
\[
\begin{tabular}{l | c}
Impurity & Strength \\ \hline\hline
$\sigma(r)=-2e^{-r^2/2}$ & $-2.11$ \\
$\sigma(r)=-4e^{-r^2}$ & $-1.82$ \\
$\sigma(r)=-8e^{-2r^2}$ & $-1.74$\\
$\sigma(r)=-16e^{-4r^2}$ & $-1.71$\\
\end{tabular}
\]
As the impurity becomes more like a delta function, the point charge approaches the value of the corresponding vortex. The derivation of the asymptotic 1-vortex metric goes through in almost exactly the same way as the asymptotic 2-vortex metric in \cite{MS03}, giving
\begin{equation}\label{asympmetric}
\pi\left(1-2qq^\prime K_0(\rho)\right)(d\rho^2+\rho^2d\theta^2),
\end{equation}
where $q$, $q^\prime$ are the point charges of the vortex and the impurity respectively. Furthermore, the impurity is located at the origin, and the vortex is at $\rho {\rm e}^{i\theta}$ for $\rho \gg 0.$

We can calculate exactly the difference in volume between \eqref{impmetric} and the metric on the impurity-free 1-vortex moduli space $\pi dz_1d\bar{z}_1$, which \eqref{impmetric} approaches asymptotically.  The exponential decay of $K_0$ as $\rho\to\infty$ and the form of \eqref{impmetric} imply that $b$ also decays exponentially, and integrating in polar coordinates easily gives this difference in volume to be $\pi\int\sigma\, d^2x$.

\subsection{Delta-function impurities}\label{deltaimp}
It is natural to consider what happens in the limit where $\sigma$ approaches a delta-function.  Unfortunately this introduces the square of a delta-function into the Lagrangian \eqref{mag}, which is not defined.  However, it does make sense to replace $\sigma$ with a delta-function in \eqref{modbog2}, and we can consider this to be a limit of impurities for which there is a Lagrangian description.

\begin{figure}[h]
\begin{center}
\includegraphics[scale=0.87]{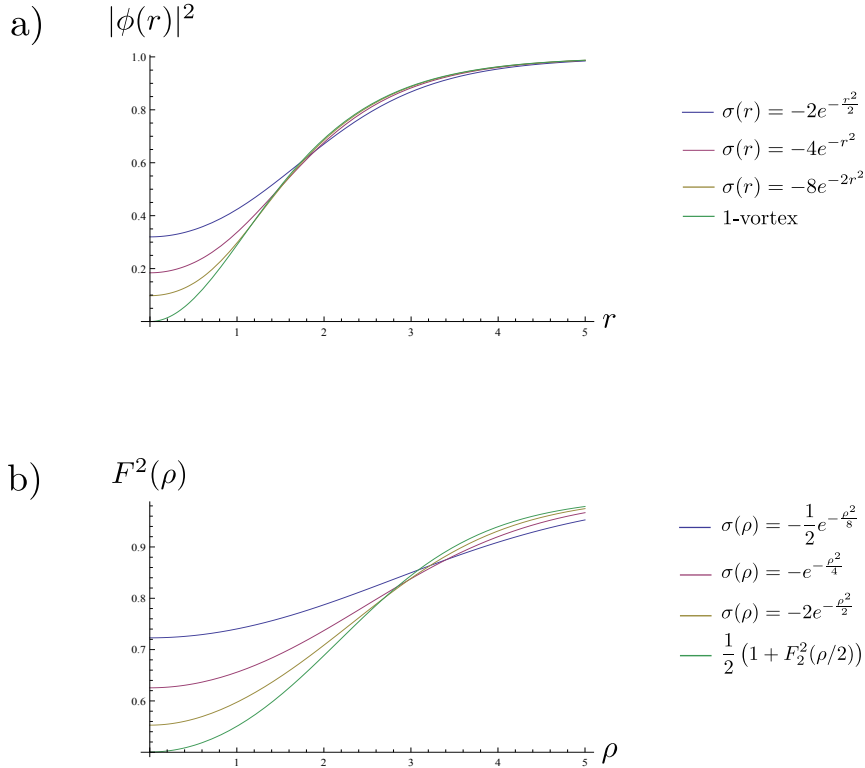}
\footnotesize{\caption{a) Higgs field profiles for vacua converging to the 1-vortex profile in the delta-function limit.  b) Moduli space metric profiles for 1-vortices in the presence of impurities converging to a shifted version of the impurity-free 2-vortex metric in the delta-function limit.\label{deltalimit}}}
\end{center} 
\end{figure}

Suppose we replace $\sigma$ by a delta-function of the form $-4\pi\alpha\delta(z)$ where $\alpha\in \mathbf{N}$, and we look for solutions with winding number $N$.  In this case \eqref{Taubes} becomes the impurity-free Taubes equation for $N+\alpha$ vortices with $\alpha$ vortices constrained to lie at $z=0$.  This corresponds to applying the singular gauge transformation $\phi\to e^{-i\alpha\theta}\phi$ to this $(N+\alpha)$-vortex solution, so the winding number of $\phi$ at infinity is still $N$.  In a formal sense, we could solve \eqref{modbog1} and \eqref{modbog2} by taking any $(N+\alpha)$-vortex solution with a single vortex at the origin and applying this singular gauge transformation, but this would not correspond to a limit of solutions to Bogomolny equations with finite impurities.  Numerical investigations suggest that the correct moduli space of solutions with winding number $N$ is the submanifold of the impurity-free $(N+\alpha)$-vortex moduli space with $\alpha$ vortices fixed at the origin.  Fig.~\ref{deltalimit} a) shows vacua with winding number zero converging to the 1-vortex solution in the delta-function limit.  

The global gauge transformation relating an $N$-vortex $(\phi,A_i)$ in the presence of an $\alpha$-impurity to an impurity-free $(N+\alpha)$-vortex $(\phi^\prime,A_i^\prime)$ is the same for any vortex configuration, which means that the corresponding map between the moduli space of $N$-vortices in the presence of an $\alpha$-impurity and the moduli space of $(N+\alpha)$-vortices with $\alpha$ vortices constrained to lie at the origin is a diffeomorphism.  Gauss's law is unchanged by the impurity, so the infinitesimal variations of the fields orthogonal to gauge orbits are unchanged by this diffeomorphism,  implying that it is an isometry.  This gives the important result that the metric on the moduli space of $N$ vortices in the presence of delta-function impurities with $\alpha\in \mathbf{N}$ is just the restriction of the usual impurity-free $(N+\alpha)$-vortex metric to the submanifold of solutions where $\alpha$ vortices are fixed at the origin.  As one would expect, the numerics suggest that the moduli space metric for finite impurities converges to the metric on this submanifold in the delta-function limit.   In particular, the metric for a 1-vortex moving in the presence of an impurity $\sigma=-4\pi\delta(z)$ should be related to the impurity-free 2-vortex metric by a simple shift of coordinates.  The impurity-free metric for two vortices at positions $z_1=Z+W$ and $z_2=Z-W$ has the form
\[
ds^2=2\pi dZ d\bar{Z}+2\pi F_2^2(|W|)\,dW d\bar{W},
\]
where $F_2$ is the centred 2-vortex metric first calculated numerically by Samols.  The submanifold defined by the constraint $z_2=0$ therefore has metric
\begin{equation}\label{1vimpmetric}
\frac{\pi}{2}\left(1+F^2_2(\rho/2)\right)(d\rho^2+\rho^2 d\theta^2),
\end{equation}
where $z_1=\rho e^{i\theta}$ as before.  Fig.~\ref{deltalimit} b) shows moduli space metrics for 1-vortices in the presence of impurities converging to \eqref{1vimpmetric}, as expected.

We can also find numerical solutions in the case where $\alpha$ is any positive real number by solving
\[
\nabla^2 \Phi+1-|z|^\alpha\prod^N_{r=1}|z-z_r|^2 e^\Phi=0.
\]
with the same boundary conditions as before.  Near the origin the Higgs field of the solution will vanish to order $\alpha$, and the interpretation is that $\alpha$ vortices are pinned at the origin.  The next section shows that we can find explicit static solutions and metrics for all $\alpha>0$ if we move to a hyperbolic space background.

\section{Hyperbolic vortices with delta-function impurities}\label{hypvordelta}
It is straightforward to generalise \eqref{mag} to a spacetime of the form $X\times \mathbf{R}$, where $X$ is an arbitrary Riemann surface.  Locally we choose isothermal coordinates so that the metric on $X$ is of the form $ds^2=\Omega(x_1,x_2)(dx_1^2+dx_2^2)$.  We can still apply a Bogomolny argument to obtain first-order vortex equations, and in this coordinate patch they are:
\begin{align}
D_1\phi+iD_2\phi=0,\label{locbog1}\\ 
B-\frac{\Omega}{2}(1+\sigma-\vert\phi\vert^2)=0,\label{locbog2}
\end{align}
and \eqref{Taubes} generalises to 
\begin{equation}\label{genmodTaubes}
\nabla^2 f+\Omega(1+\sigma-e^f)=4\pi\sum^N_{r=1}\delta^{2}(z-z_r).
\end{equation}
We will work on the disc model of hyperbolic space with curvature $-1/2$, which means choosing a single global chart for which
\begin{equation}\label{freemetric}
\Omega(z,\bar{z})=\frac{8}{(1-|z|^2)^2},
\end{equation}
where $|z|<1$.  Witten showed some time ago \cite{W78} that the impurity-free vortex equations on this background are integrable.  If we make the substitution $f=2g+2\log \frac{1}{2}(1-|z|^2)$, then \eqref{genmodTaubes} becomes
\begin{equation}\label{hypimp}
\nabla^2g+\frac{1}{2}\Omega\sigma-e^{2g}=2\pi\sum^N_{r=1}\delta^2(z-z_r).
\end{equation}
When $\sigma=0$, this is Liouville's equation with sources, whose solution is
\[
g=-\log\frac{1}{2}\left(1-|h|^2\right)+\frac{1}{2}\log\left\lvert\frac{dh}{dz}\right\rvert^2,
\]
where $h(z)$ is an analytic function of the form
\begin{equation}\label{Blaschke}
h(z)=z\prod^{N}_{i=1}\left(\frac{z-\beta_i}{1-\bar{\beta_i}z}\right).
\end{equation}
The $\beta_i$ are complex numbers in the unit disc chosen so that $\frac{dh}{dz}$ vanishes at the vortex positions $z_r$.  The corresponding Higgs and gauge fields are
\[
\phi=\frac{1-\vert z\vert^2}{1-\vert h \vert^2}\frac{dh}{dz}\textrm{ and }A_{\overline{z}}=-i\frac{\partial}{\partial \bar{z}}\log \left(\frac{1-\vert z\vert^2}{1-\vert h \vert^2}\right),
\]
where $A_{\bar{z}}=\frac{1}{2}(A_1+iA_2)$.  

If we choose $\sigma(z)=-4\pi\alpha\Omega(0)^{-1}\delta(z)$ for any positive $\alpha$, then the equations \eqref{locbog1} and \eqref{locbog2} can still be solved by the same rational map ansatz which constrains $\alpha$ of the vortices to lie at the origin.  We let

\begin{equation}
h(z)=z^{\alpha+1}\prod^{N}_{i=1}\left(\frac{z-\beta_i}{1-\bar{\beta_i}z}\right) = z^{\alpha+1} {\tilde h}(z).
\end{equation}
The corresponding Higgs field is
\[
\phi_{{\rm vortices}} = \frac{1-|z|^2}{1-|z|^{2\alpha+2} |{\tilde f}|^2} \left((\alpha+1)z^{\alpha}{\tilde h} +z^{\alpha+1}\frac{d{\tilde h}}{dz}\right),
\]
so the $\beta_i$ must be chosen so that the $N$ zeroes of $z\frac{d{\tilde h}}{dz}+(\alpha+1){\tilde h}$ lie at the vortex positions $z_r$.  This Higgs field does not quite correspond to $N$ vortices in the presence of an impurity of strength $\alpha$, because it has winding number $N+\alpha$ and is multi-valued for non-integer $\alpha$.  As discussed in section \ref{deltaimp}, one must apply a singular gauge transformation of the form $\phi\to \frac{|z|^\alpha}{z^\alpha}\phi$. The resulting (single-valued) fields are
\begin{equation}\label{arb}
\phi=|z|^\alpha\left(\frac{1-|z|^2}{1-|z|^{2(\alpha+1)}|{\tilde h}|^2}\right)\left((\alpha+1){\tilde h}+z\frac{d{\tilde h}}{dz}\right)\textrm{ and }A_{\bar{z}}=-i\partial_{\bar{z}}\log \phi.
\end{equation}

We can make use of previous exact results discovered in \cite{KS09} to calculate exact moduli space metrics for vortices in the presence of impurities. The focus of \cite{KS09} is on submanifolds of the moduli space corresponding to cyclically symmetric configurations, for which ${\tilde h}$ takes the form
\begin{equation}\label{cycansatz}
{\tilde h(z)}=\frac{z^N-a^N}{1-\bar{a}^Nz^N}.
\end{equation}
In this interpretation the vortex solution corresponding to \eqref{cycansatz} consists of $N$ vortices arranged in a regular $N$-gon around $\alpha$ coincident vortices at the origin.  If $N>\alpha$ then these configurations exhaust the space $\Sigma_{N,\alpha}$ of $C_N$-symmetric $(N+\alpha)$-vortices, which is a totally geodesic submanifold of the full moduli space. The metric on $\Sigma_{N,\alpha}$ was calculated to be \cite{KS09}:
\begin{equation}\label{cycmetric}
ds^2=\frac{4\pi N^3|z_1|^{2N-2}}{(1-|z_1|^{2N})^2}\left(1+\frac{2N(1+|z_1|^{2N})}{\sqrt{(\alpha+1)^2(1-|z_1|^{2N})^2+4N^2|z_1|^{2N}}}\right)dz_1d\bar{z}_1,
\end{equation}
where $z_1$ is the position of one of the outer $N$ vortices.  If we interpret \eqref{cycansatz} as giving an $N$-vortex in the presence of a delta-function impurity of strength $\alpha$, then $\alpha$ can be any real number and we no longer have the restriction that $N>\alpha$, but the derivation is unchanged and moduli space metric is still given by \eqref{cycmetric}.

\begin{figure}[h]
\begin{center}
\includegraphics[scale=0.45]{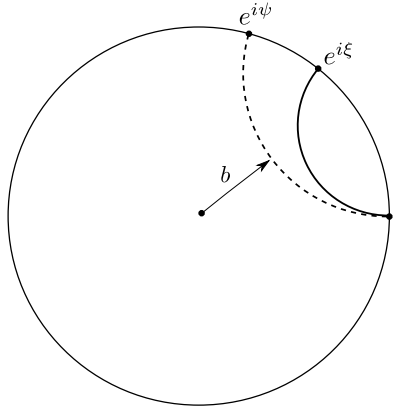}
\caption{Diagram illustrating the definition of impact parameter and scattering angle for a vortex scattering off an impurity in hyperbolic space.\label{saipdefn}}
\end{center}
\end{figure}

\begin{figure}[h]
\begin{center}
\includegraphics[scale=0.75]{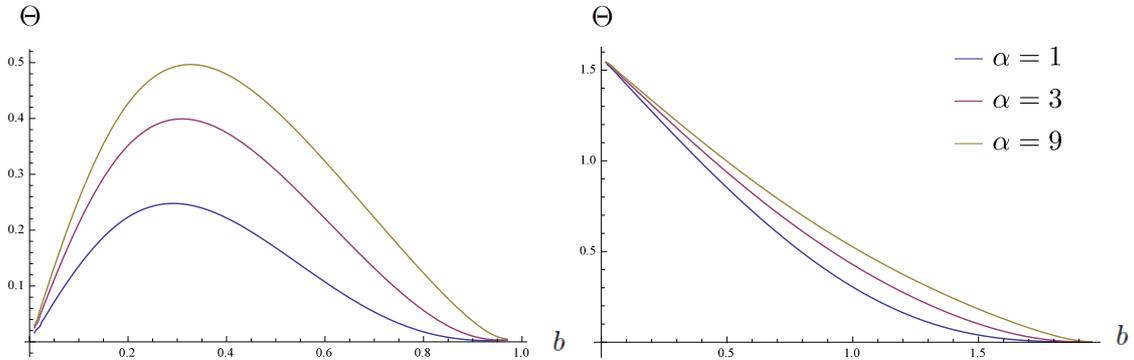}
\caption{Scattering angle versus impact parameter for various values of the impurity strength.\label{saip}}
\end{center}
\end{figure}

Fig.~\ref{saip} gives a plot of scattering angle versus Euclidean impact parameter for $N=1, 2$ and various values of $\alpha$ in the disc model.  For $N=2$ scattering angle and impact parameter in are defined as in \cite{KS09}, while Fig.~\ref{saipdefn} shows how scattering angle and impact parameter are defined for $N=1$: given a vortex trajectory, the impact parameter $b$ is defined to be the distance between the impurity and the closest point to it on the geodesic of hyperbolic space which makes second order contact with the incoming end of the trajectory.  If this geodesic starts at $1$ and ends at $e^{i\psi}$, and the outgoing end of the vortex trajectory is at $e^{i\xi}$, then the scattering angle is defined to be $\Theta=\psi-\xi$.  Fig.~\ref{saip} shows that, regardless of the strength of the impurity, the incoming vortex passes straight through the impurity if the impact parameter is zero.  As one would expect, vortices are deflected more strongly for greater impurity strengths, with the deflection tending to zero as the Euclidean impact parameter tends to $1$.

The limit $\alpha\to \infty$ corresponds to a fixed `bag' of hyperbolic vortices at the origin of the type considered in \cite{S12}.  The Higgs field is very close to zero inside the bag, whose radius grows with $\alpha$, and very close to $1$ outside the bag.  The thickness of the bag's surface depends only on the Higgs mass. We can see this bag structure in the metric $f^\alpha_1(|z_1|)\,dzd\bar{z}$ for a 1-vortex in the presence of an impurity of strength $\alpha$.  The disc model coordinate $z_1$ is not convenient for analysing the large $\alpha$ limit, because it has finite range.  If we change coordinates $|z_1|=\tanh(r/2^{3/2})$, then the hyperbolic plane metric with conformal factor \eqref{freemetric} becomes
\[
ds^2=dr^2+2\sinh^2(r/\sqrt{2})d\theta^2,
\]
and $r$ has infinite range.  The radius of the bag is given approximately by $R=\sqrt{2}\log (2\alpha)$.  If we scale coordinates $r^\prime =r/R$ to keep this radius at the same position as $\alpha$ changes, then it is easy to check that as $\alpha\to \infty$, $f^\alpha_1(\tanh(Rr^\prime/2^{3/2}))$ tends pointwise to the step function
\[
f_\infty(r^\prime)=\begin{cases}  4\pi & \text{ if $r^\prime<1$}\\
12\pi & \text{ if $r^\prime>1$}
\end{cases}
\]
This shows that a slow-moving vortex moves along the geodesics of hyperbolic space both inside and outside the bag, but with effective inertial masses $\pi/2$ inside the bag and $3\pi/2$ outside.
\section{Vortices on the 2-sphere with impurities near the Bradlow limit}\label{sphvorimp}
In this subsection we adapt the results of Baptista and Manton on vortices on the 2-sphere near the Bradlow limit to the inclusion of delta-function magnetic impurities.  Since the background is a closed manifold, we think of $\phi$ as a section of a complex line bundle $E\to S^2\cong \mathbf{CP}^1$ equipped with a Hermitian metric $h$.  The bundle $E\to S^2$ can now have non-trivial topology and we identify its first Chern class with the topological charge $N$.  The gauge potentials are local 1-form representatives of an $h$-compatible connection $D$ on $E$, and the magnetic field is identified with the curvature $F\in \Omega^2(X,\mathbf{R})$ of $D$.  The Bogomolny equations can be written in covariant form as
\begin{align}
D^{0,1}\phi=0,\label{globog1}\\
F-\frac{1}{2}(1+\sigma-|\phi|^2_h)\text{vol}_X=0,\label{globog2}
\end{align}
where the impurity $\sigma$ is some smooth real-valued function on $S^2$.  

We will take the usual atlas on $\mathbf{CP}^1$ consisting of the two open sets $U_1=\mathbf{CP}^1\setminus \{ [0,1] \}$ and $U_2=\mathbf{CP}^1\setminus \{ [1,0] \}$ and the charts $\varphi_i:U_i\to \mathbf{C}$ with transition function $\varphi_1\circ\varphi_2^{-1}(z)=1/z$.  We will consider spheres of varying radius with metric defined by $g_R=R^2\times$ (standard round sphere metric).  The line bundle $\pi:E\to S^2$ is defined by the transition functions $g_{ij}:U_i\cap U_j\to U(1)$ where
\[
g_{21}\circ\varphi_2^{-1}(z)=(z/|z|)^N,\quad g_{12}=1/g_{21},\quad g_{11}=g_{22}=1.
\]
It is straightforward to check that these transition functions satisfy the cocycle conditions and that the corresponding bundle has degree $N$.  If $\psi_i:\pi^{-1}(U_i)\to U_i\times \mathbf{C}$ are the associated trivialisations, then one can define a metric $h$ by setting $|\psi^{-1}_i(p,y)|^2_h=|y|^2$.  This is the bundle and metric we shall take throughout this subsection.

Integrating \eqref{globog2} over the 2-sphere gives an obstruction to the existence of solutions to the Bogomolny equations (known as the Bradlow bound when $\sigma=0$):
\begin{equation}\label{modBrad}
R^2+\frac{1}{4\pi}\int \sigma\,\text{vol}_R\geq N.
\end{equation}
Suppose we have a delta-function impurity $\sigma$ defined by $\sigma(\phi_1(z))=-\frac{\pi\alpha}{R^2}\,\delta^2(z)$ on $U_1$ and $\sigma(\phi_2(z))=0$ on $U_2$, so that $\int \sigma\,\text{vol}_R=-4\pi\alpha$.  We can explicitly solve the Bogomolny equations at the Bradlow limit $R^2=N+\alpha$ for this choice of impurity.  The Higgs field must vanish everywhere if the bound is saturated, and one can check that the gauge potentials $A_i\in\Omega^1(U_i,\mathbf{R})$ defined by
\begin{align*}
A_1&=\phi^*_1A+\frac{i\alpha}{2|z|^2}(\overline{z}dz-zd\overline{z}),\\
A_2&=\phi^*_2A,
\end{align*}
where
\[
A=-i\frac{N+\alpha}{2(1+|z|^2)}(\bar{z}dz-zd\bar{z})
\]
give a connection $D_{N+\alpha}$ on $E$ with $F_{N+\alpha}=\frac{1}{2}(1+\sigma)\text{vol}_{\sqrt{N+\alpha}}$ as required.

Now we move away from the Bradlow limit, taking $R^2$ to be slightly greater than $N+\alpha$.  We shall make similar assumptions to Baptista and Manton that $D\approx D_{N+\alpha}$ for these vortices, and that $\phi$ satisfies the conditions:
\begin{enumerate}
\item $D_{N+\alpha}^{0,1}\phi=0$\label{cond1}
\item $\int_{\mathbf{CP}^1}\left(F_{N+\alpha}-\frac{1}{2}(1+\sigma-|\phi|^2_h)\text{vol}_R\right)=0$\label{cond2}
\end{enumerate}
We shall work on the coordinate patch $U_1$ and take a representative $\phi_1=\phi\circ \varphi_1^{-1}$ of $\phi$.  In these coordinates condition \ref{cond1} becomes
\[
\frac{\partial\phi_1}{\partial\bar{z}}=z\left(-\frac{\alpha+N}{2(1+|z|^2)}+\frac{\alpha}{2|z|^2}\right)\phi_1,
\]
which has general solution 
\[
\phi_1=\frac{f(z)|z|^{\alpha}}{(1+|z|^2)^{(N+\alpha)/2}},
\]
where $f$ is holomorphic on $\mathbf{C}$.  This must be extensible to a solution $\phi_2(z)=g_{12}(z)\phi_1(\frac{1}{z})$ of condition \ref{cond1} on $U_2$, which forces $f$ to be a polynomial in $z$ of degree $N$.  If we write ${f(z)=a_0z^N+a_1z^{N-1}+\dots+a_N}$, then condition \ref{cond2} becomes
\begin{align}
4\pi(R^2-N-\alpha)=\int_{\mathbf{CP}^1} |\phi|_h^2\,\text{vol}_R &=\int_{\mathbf{C}}|\phi_1|^2\frac{2iR^2}{(1+|z|^2)^2}dz\wedge d\bar{z}\\&=\sum^N_{k=0}|a_{k}|^2\int_\mathbf{C}\frac{|z|^{2(N-k+\alpha)}2iR^2}{(1+|z|^2)^{N+\alpha+2}}dz\wedge d\bar{z}\\
&=4\pi R^2\sum^N_{k=0}|a_k|^2\frac{\Gamma(k+\alpha+1)(N-k)!}{\Gamma(N+\alpha+2)}
\end{align}
Just as for impurity-free vortices, there is a bijection between the space of solutions to conditions \ref{cond1}, \ref{cond2} and $S^{2N+1}$ given by
\begin{equation}\label{ident}
\phi_1\to \left(1-\frac{N+\alpha}{R^2}\right)^{-1/2}\left(\dots,\left(\frac{\Gamma(k+\alpha+1)(N-k)!}{\Gamma(N+\alpha+2)}\right)^{1/2}a_k,\dots \right)_{0\leq k\leq N}
\end{equation}
\begin{figure}[!h]
\begin{center}
\begin{subfigure}{0.48\textwidth}
\includegraphics[scale=0.87]{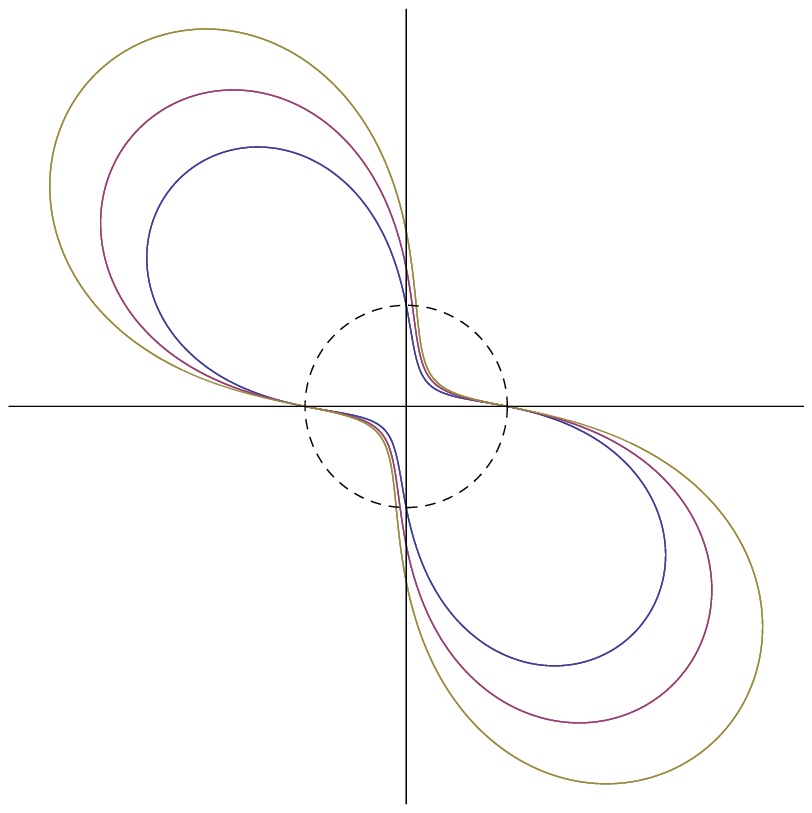}
\caption{Initial positions $z_1=1$, $z_2=-1$ and velocities $\dot{z}_1=-1+0.2i$, $\dot{z}_2=1-0.2i$.}
\end{subfigure}\quad
\begin{subfigure}{0.48\textwidth}
\includegraphics[scale=0.87]{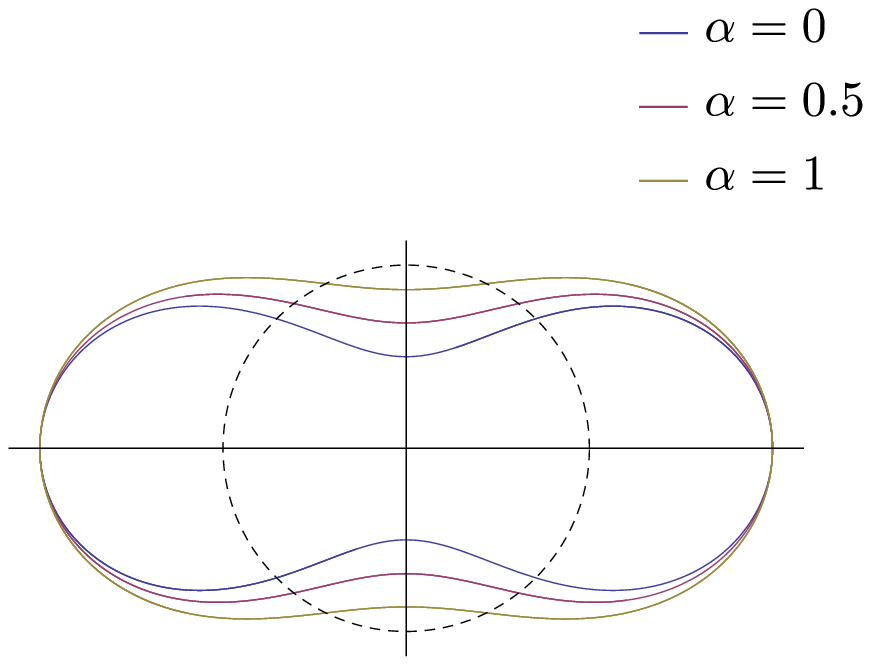}
\caption{Initial positions $z_1=2$, $z_2=2$ and velocities $\dot{z}_1=2$, $\dot{z}_2=2$.}
\end{subfigure}
\footnotesize{\caption{Geodesics corresponding to slow-motion scattering of two vortices off a delta-function impurity at the south pole.  The southern hemisphere has been stereographically projected to the interior of the unit disc.\label{2vordelta}}}
\end{center}
\end{figure}
The rest of the derivation of the moduli space metric in \cite{BM03} goes through unchanged; the only difference is the sphere identification \eqref{ident}.  The fact that we have fixed the gauge field means that the remaining gauge freedom is multiplication of $\phi$ by a constant phase, so identifying gauge-equivalent points in $S^{2N+1}$ corresponds to the usual $U(1)$-principal bundle $S^{2N+1}\to \mathbf{CP}^N$. The geodesic approximation can be implemented in just the same way as in the impurity-free case, and the metric on $\mathbf{CP}^N$ is again the Fubini-Study metric. In particular, the 1-vortex metric is
\begin{equation}\label{1vorimpmet}
\left(2\pi-N-\alpha\right)\frac{1+\alpha}{\left(1+\alpha+|z|^2\right)^2} dzd\bar{z}.
\end{equation}
It is easy to see from this that the vortex is pushed away from the impurity for $\alpha>0$, as one would expect from the results of the previous sections.  The same effect is visible for higher charge vortices: Fig.~\ref{2vordelta} shows two vortices scattering off the impurity for different values of $\alpha$.

\section{Conclusion}
In this paper, we investigated the effect of magnetic impurities on 
vortex dynamics in various models, based on the Lagrangian suggested in 
\cite{TW14}.  In flat space, our numerical results indicated that a moduli space of multi-vortex solutions in the presence of magnetic impurities exists.  For low charges we calculated the metrics on these 
spaces numerically. Our numerics also suggested that the static solutions 
have a well-defined limit as the impurity tends to a delta-function.

On the hyperbolic plane we found explicit solutions and moduli space 
metrics when the impurity is a delta function. This gives a new 
interpretation of the metrics on submanifolds of the moduli space discussed in \cite{KS09}, which 
parametrizes $n$ vortices at the vertices of a regular $n$-gon with $m$ vortices at the origin. The $m$ 
vortices are reinterpreted as the effect of a delta-function impurity of strength $m$. This new interpretation gives physical meaning to the submanifolds of the moduli space given in \cite{KS09} which are not totally geodesic.
We went on to numerically derive scattering angle/impact parameter plots for one and two 
vortices in the presence of an impurity at the origin. 

We treated the moduli space of vortices on the $2$-sphere near the Bradlow limit 
with a delta-function impurity in much the same way as in \cite{BM03}, showing that its moduli space metric can again be approximated by a multiple of the 
Fubini-Study metric.  An interesting future direction would be to investigate the effect of magnetic impurities on compact hyperbolic surfaces, since \cite{MM15} recently gave new analytic solutions for vortices on such surfaces, following earlier work in \cite{MR10}.

While we only focussed on the slow-motion dynamics of vortices in relativistic theories, there 
does exist an interesting first-order system of vortex dynamics which was introduced 
by Manton in \cite{M97}. This Schr\"odinger-Chern-Simons 
dynamics allows for a moduli space description close to critical coupling $\lambda=1$. The equations of 
motion are governed by the K\"ahler form on the moduli space. The 
moduli space approximation predicts that two vortices go around each 
other at constant speed with depends on $|\lambda -1|$, and this has been 
studied for more than two vortices in \cite{RS04}. The validity 
of this approximation has been verified numerically in 
\cite{KS05} and discussed analytically in 
\cite{DS09}. It would be interesting to derive the 
effects of impurities on this type of vortex dynamics.

\section*{Acknowledgments}
AHC thanks Felipe Contatto, Derek Harland, and Richard Ward for useful discussions. SK would like to thank Mareike Haberichter for interesting discussions. AM acknowledges SMSAS at the University of Kent for a PhD studentship. AHC was financially supported by an STFC studentship.

\bibliography{references}{}

\begin{thebibliography}{10}

\bibitem{TW14}
D.~Tong and K.~Wong, ``{Vortices and Impurities},'' {\em JHEP}, vol.~1401,
  p.~090, 2014.

\bibitem{Zhang}
R.~Zhang and H.~Li, ``Sharp existence theorems for multiple vortices induced
  from magnetic impurities,'' {\em Nonlinear Analysis}, vol.~115, pp.~117--129,
  2015.

\bibitem{Han:2015yua}
X.~Han and Y.~Yang, ``{Topologically Stratified Energy Minimizers in a Product
  Abelian Field Theory},'' {\em Nucl. Phys.}, vol.~B898, pp.~605--626, 2015.

\bibitem{Han:2015tga}
X.~Han and Y.~Yang, ``{Magnetic Impurity Inspired Abelian Higgs Vortices},''
  2015.

\bibitem{W78}
E.~Witten, ``Some exact multi-pseudoparticle solutions of classical yang-mills
  theory,'' {\em Phys. Rev. Lett.}, p.~121, 1976.

\bibitem{S92}
I.~A.~B. Strachan, ``{Low velocity scattering of vortices in a modified Abelian
  Higgs model},'' {\em J. Math. Phys.}, vol.~33, p.~102, 1992.

\bibitem{BM03}
J.~M. Baptista and N.~S. Manton, ``{The Dynamics of vortices on S**2 near the
  Bradlow limit},'' {\em J. Math. Phys.}, vol.~44, p.~3495, 2003.

\bibitem{JT80}
A.~Jaffe and C.~Taubes, {\em Vortices and monopoles}.
\newblock Birkh\"auser, 1980.

\bibitem{M81}
N.~Manton, ``{A Remark on the Scattering of BPS Monopoles},'' {\em Phys.
  Lett.}, vol.~B110, p.~54, 1982.

\bibitem{Stu94vor}
D.~Stuart, ``Dynamics of abelian higgs vortices in the near bogomolny regime,''
  {\em Commun. Math. Phys.}, vol.~159, no.~1, p.~51, 1994.

\bibitem{Stu94mon}
D.~Stuart, ``{The Geodesic approximation for the Yang-Mills Higgs equations},''
  {\em Commun. Math. Phys.}, vol.~166, p.~149, 1994.

\bibitem{MS04}
N.~S. Manton and P.~M. Sutcliffe, {\em Topological Solitons}.
\newblock Cambridge University Press, 2004.

\bibitem{S91}
T.~Samols, ``{Vortex scattering},'' {\em Commun. Math. Phys.}, vol.~145,
  p.~149, 1992.

\bibitem{W79a}
E.~J. Weinberg, ``Multivortex solutions of the ginzburg-landau equations,''
  {\em Phys. Rev. D}, vol.~19, p.~3008, May 1979.

\bibitem{MS03}
N.~Manton and J.~Speight, ``{Asymptotic interactions of critically coupled
  vortices},'' {\em Commun. Math. Phys.}, vol.~236, p.~535, 2003.

\bibitem{Sp96}
J.~M. Speight, ``{Static intervortex forces},'' {\em Phys. Rev.}, vol.~D55,
  p.~3830, 1997.

\bibitem{deVega:1976mi}
H.~J. de~Vega and F.~A. Schaposnik, ``{A Classical Vortex Solution of the
  Abelian Higgs Model},'' {\em Phys. Rev.}, vol.~D14, pp.~1100--1106, 1976.

\bibitem{T02}
D.~Tong, ``{NS5-branes, T duality and world sheet instantons},'' {\em JHEP},
  vol.~0207, p.~013, 2002.

\bibitem{Ohashi:2015yta}
K.~Ohashi, ``{Small Winding-Number Expansion: Vortex Solutions at Critical
  Coupling},'' {\em JHEP}, vol.~11, p.~073, 2015.

\bibitem{KS09}
S.~Krusch and J.~M. Speight, ``{Exact moduli space metrics for hyperbolic
  vortices},'' {\em J.Math.Phys.}, vol.~51, p.~022304, 2010.

\bibitem{S12}
P.~M. Sutcliffe, ``{Hyperbolic vortices with large magnetic flux},'' {\em Phys.
  Rev.}, vol.~D85, p.~125015, 2012.

\bibitem{MM15}
R.~Maldonado and N.~S. Manton, ``{Analytic vortex solutions on compact
  hyperbolic surfaces},'' {\em J. Phys.}, vol.~A48, no.~24, p.~245403, 2015.

\bibitem{MR10}
N.~S. Manton and N.~A. Rink, ``{Vortices on Hyperbolic Surfaces},'' {\em J.
  Phys.}, vol.~A43, p.~434024, 2010.

\bibitem{M97}
N.~S. Manton, ``{First order vortex dynamics},'' {\em Annals Phys.}, vol.~256,
  pp.~114--131, 1997.

\bibitem{RS04}
N.~M. Romao and J.~M. Speight, ``{Slow Schr\"odinger dynamics of gauged
  vortices},'' {\em Nonlinearity}, vol.~17, pp.~1337--1355, 2004.

\bibitem{KS05}
S.~Krusch and P.~Sutcliffe, ``{Schr\"odinger-Chern-Simons vortex dynamics},''
  {\em Nonlinearity}, vol.~19, pp.~1515--1534, 2006.

\bibitem{DS09}
S.~Demoulini and D.~Stuart, ``{Adiabatic limit and the slow motion of vortices
  in a Chern-Simons-Schr\"odinger system},'' {\em Commun. Math. Phys.},
  vol.~290, pp.~597--632, 2009.

\end{thebibliography}
\bibliographystyle{ieeetr}
\end{document}